\RequirePackage{iftex}
\ifPDFTeX
\pdfoutput=1
\fi
\documentclass[11pt]{article}

\usepackage[utf8]{inputenc}
\usepackage[T1]{fontenc}
\usepackage{iftex}
\usepackage{amsmath, amssymb}
\usepackage{array}
\usepackage{graphicx}
\usepackage{booktabs}
\usepackage{microtype}
\ifPDFTeX
\usepackage{hyperref}
\else
\usepackage[xetex]{hyperref}
\fi
\usepackage[margin=1in]{geometry}
\usepackage{cite}
\usepackage{xcolor}
\usepackage{tikz}
\usepackage{pgfplots}
\usepgfplotslibrary{fillbetween,groupplots}
\pgfplotsset{compat=1.18}
\hypersetup{
  colorlinks=true,
  linkcolor=blue,
  citecolor=blue,
  urlcolor=blue,
}

\definecolor{plotblue}{RGB}{30,144,255}
\definecolor{plotcyan}{RGB}{0,230,230}
\definecolor{plotgreen}{RGB}{0,220,100}
\definecolor{plotyellow}{RGB}{255,220,50}
\definecolor{plotred}{RGB}{255,80,80}

\pgfplotsset{
  darktheme/.style={
    axis background/.style={fill=white},
    axis line style={color=black},
    tick style={color=black},
    ticklabel style={color=black},
    label style={color=black},
    title style={color=black},
    grid=both,
    grid style={draw=black!10, thin},
    major grid style={draw=black!20},
    legend style={
      fill=white,
      draw=black!20,
      text=black,
      font=\footnotesize,
      cells={anchor=west},
    },
    every axis plot/.append style={very thick},
  }
}

\title{Modernizing Amdahl's Law\\\large\parbox{0.9\textwidth}{\centering How AI Scaling Laws Shape Computer Architecture}}
\author{Chien-Ping Lu}
\date{}

\begin{document}

\maketitle

\begin{abstract}
\noindent
Classical Amdahl's Law conceptualized the limit of speedup for an era of
fixed serial--parallel decomposition and homogeneous replication.
Modern heterogeneous systems need a different conceptual framework:
constrained resources must be allocated across heterogeneous hardware
while workloads themselves change, with some stages becoming effectively
bounded and others continuing to absorb additional effective compute.
This paper reformulates Amdahl's Law around that shift. We replace
processor count with an allocation variable,
replace the classical parallel fraction with a \emph{value-scalable
fraction}, and model specialization by a relative efficiency ratio
between dedicated and programmable compute. The resulting objective
yields a finite collapse threshold. For a specialized efficiency ratio
$R$, there is a critical scalable fraction $S_c = 1 - 1/R$ beyond which
the optimal allocation to specialization becomes zero. Equivalently, for
a given scalable fraction $S$, the minimum efficiency ratio required to
justify specialization is $R_c = 1/(1-S)$. Thus, as value-scalable
workload grows, over-customization faces a rising bar. The point is not
that one hardware class simply defeats another, but that architecture
must preserve a sufficiently programmable substrate against a moving
frontier of work whose marginal gains keep scaling.  In practice, that
frontier is often sustained by software- and model-driven efficiency
doublings rather than by fixed-function redesign alone.  The model helps
explain the migration of value-producing work toward learned late-stage
computation and the shared design pressure that is making both GPUs and
AI accelerators more programmable.
\end{abstract}

\section{Introduction}
\label{sec:introduction}

Amdahl's Law~\cite{amdahl1967} has long served as a foundational model
for reasoning about performance scaling:
\begin{equation}
  \text{Speedup} = \frac{1}{(1-P) + \dfrac{P}{N}}
  \label{eq:classical}
\end{equation}
where $P$ denotes the fraction of work assumed to be parallelizable and
$N$ denotes the number of processors.

This formulation assumes:
\begin{itemize}
  \item homogeneous compute units,
  \item replication-based scaling, and
  \item a fixed decomposition between serial and parallel components.
\end{itemize}

These assumptions no longer reflect modern systems.  Modern platforms
are heterogeneous: they mix programmable compute, dedicated units,
tensor engines, mixed precision arithmetic, and layered memory systems
within a single device.  In that setting, ``speedup'' and ``core count''
are no longer sufficient physical descriptions.  Even the term ``core''
now covers a wide range of structures, from SIMD lanes and tensor
datapaths to fixed-function assist blocks.  Hennessy and Patterson
described the present period as a ``new golden age for computer
architecture''~\cite{hennessy2019}, driven by domain-specific efficiency
optimization.  Our point is not to sort hardware into fixed dedicated
and general-purpose camps, but to ask this: if the dominant workload has
changed, then what should count as
programmable or general-purpose compute in the first place?

\section{Programmability Is a Shifting Concept}

Programmability is itself a shifting concept rather than a timeless
binary property.  CPUs are often treated as maximally programmable, yet
their cache hierarchies, branch prediction, and execution models are
already tuned to common workload structure.  Graphics processors were
once regarded as more specialized because graphics and data-parallel
programs were treated as special cases; as those workloads became
mainstream, the same substrate
came to count as programmable.  More generally, when a formerly
domain-specific workload becomes powerful enough to define the
mainstream, the architecture shaped around it begins to count as general
purpose compute for that era.  Under AI scaling, that shift continues:
terms such as KV caches, context windows, expert parallelism, and
inference-time strategies are becoming part of the architectural
vocabulary.  Readers should likewise accept that Transformer
architecture has itself become part of today's architectural vocabulary,
rather than remaining merely an application-layer detail. 
Transformer-based large language models are the clearest current example: what once
looked domain-specific is now shaping the mainstream substrate and the
language used to describe it.  CPU-era programmability was mainly about
supporting many different functions across many applications, so that
software upgrades could deliver new functionality on existing hardware.
AI-era programmability is different: a single model can already serve
many functions, so the relevant programmability is increasingly about
expanding model capability, capacity, and modes of use, allowing more
capable models to emerge on existing hardware as software and model
design improve.  The practical point is that the scaling-law relation
can remain comparatively stable even while the implementations feeding
it---algorithms, systems, software, and hardware---keep changing
underneath~\cite{lu2026}.

\begin{table}[htbp]
\centering
\small
\begin{tabular}{|
  >{\raggedright\arraybackslash}p{0.19\linewidth}|
  >{\raggedright\arraybackslash}p{0.23\linewidth}|
  >{\raggedright\arraybackslash}p{0.23\linewidth}|
  >{\raggedright\arraybackslash}p{0.23\linewidth}|}
\hline
\textbf{Aspect} & \textbf{CPU programmability}
& \textbf{Graphics programmability}
& \textbf{AI programmability} \\
\hline
What the user sees & New application functions
& Better visual quality and richer effects
& More capable models and better responses \\
\hline
Why it matters & One machine must support many tasks
& Software can replace fixed stages with richer ones
& Software can keep improving models between redesigns \\
\hline
Architectural meaning & General instruction substrate
& Shared graphics-compute substrate
& Shared tensor-memory-interconnect substrate \\
\hline
\end{tabular}
\caption{Programmability is a shifting architectural concept.  What
  counts as ``programmable'' changes with the dominant workload and with
  the kind of improvement that software can continue to unlock on
  existing hardware.}
\label{tab:programmability-eras}
\end{table}

In this paper,
``programmable''
therefore means a software-mediated substrate able to absorb the
evolving mainstream workload and the efficiency doublings that scaling
laws keep demanding~\cite{lu2026}, not a fixed notion of generality.

\section{From Classical Amdahl to Hardware Allocation}
\label{sec:limitations}

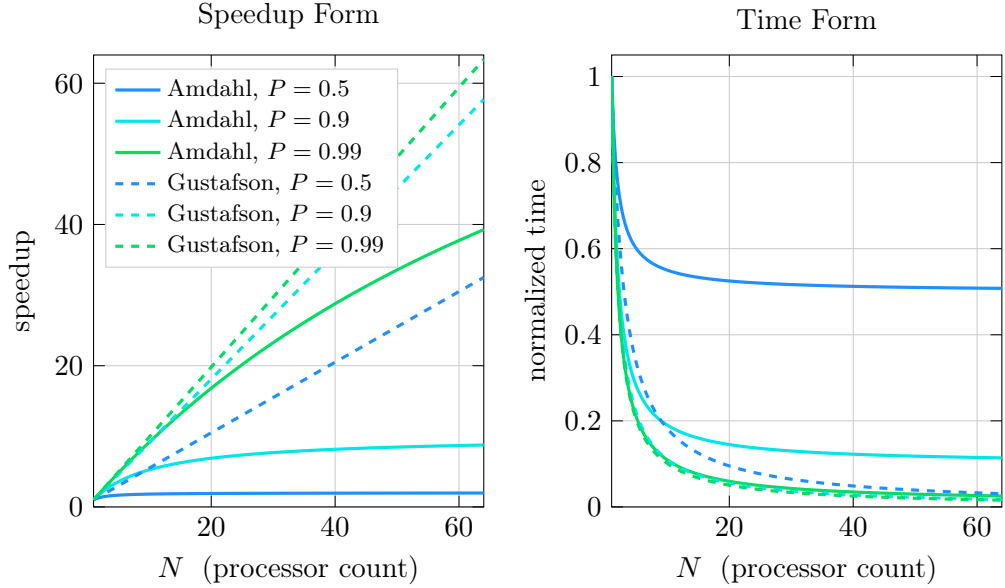
\begin{figure}[htbp]
\centering
\begin{tikzpicture}
\begin{groupplot}[
  group style={
    group size=2 by 1,
    horizontal sep=1.7cm,
  },
  darktheme,
  width=0.41\linewidth,
  height=0.46\linewidth,
  domain=1:64,
  samples=200,
  xmin=1, xmax=64,
  xlabel={$N$ \;(processor count)},
]

\nextgroupplot[
  ymin=0, ymax=64,
  ylabel={speedup},
  legend pos=north west,
  title={Speedup Form},
]

\addplot[color=plotblue]
  {1 / (0.5 + 0.5/x)};
\addlegendentry{Amdahl, $P=0.5$}

\addplot[color=plotcyan]
  {1 / (0.1 + 0.9/x)};
\addlegendentry{Amdahl, $P=0.9$}

\addplot[color=plotgreen]
  {1 / (0.01 + 0.99/x)};
\addlegendentry{Amdahl, $P=0.99$}

\addplot[color=plotblue, dashed]
  {0.5 + 0.5*x};
\addlegendentry{Gustafson, $P=0.5$}

\addplot[color=plotcyan, dashed]
  {0.1 + 0.9*x};
\addlegendentry{Gustafson, $P=0.9$}

\addplot[color=plotgreen, dashed]
  {0.01 + 0.99*x};
\addlegendentry{Gustafson, $P=0.99$}

\nextgroupplot[
  ymin=0, ymax=1.05,
  ylabel={normalized time},
  title={Time Form},
]

\addplot[color=plotblue]
  {0.5 + 0.5/x};

\addplot[color=plotcyan]
  {0.1 + 0.9/x};

\addplot[color=plotgreen]
  {0.01 + 0.99/x};

\addplot[color=plotblue, dashed]
  {1 / (0.5 + 0.5*x)};

\addplot[color=plotcyan, dashed]
  {1 / (0.1 + 0.9*x)};

\addplot[color=plotgreen, dashed]
  {1 / (0.01 + 0.99*x)};

\end{groupplot}
\end{tikzpicture}
\caption{Historical legacy of classical scaling laws.  Amdahl's law
  (solid) and Gustafson's law (dashed) shown side by side in speedup
  form and normalized-time form.  Both are expressed in terms of
  processor count~$N$ and replication-based speedup, which is the
  framing the present paper leaves behind.}
\label{fig:legacy-laws}
\end{figure}

Classical Amdahl analysis writes normalized time as
\begin{equation}
  T_A(N) = (1-P) + \frac{P}{N}
  \label{eq:amdahl-time}
\end{equation}
under a fixed decomposition between serial and parallel work.  Gustafson's
Law~\cite{gustafson1988} relaxes that fixed decomposition by allowing the
effective workload ratio to change with system scale:
\begin{equation}
  \text{Speedup}_G = (1-P) + P \cdot N.
  \label{eq:gustafson}
\end{equation}
Equivalently, in normalized-time form,
\begin{equation}
  T_G(N) = \frac{1}{(1-P) + P \cdot N}
         = \bigl(1-\tilde{P}(N)\bigr) + \frac{\tilde{P}(N)}{N},
  \label{eq:gustafson-time}
\end{equation}
with the equivalent scale-dependent fraction
\begin{equation}
  \tilde{P}(N) = \frac{N P}{(1-P) + P \cdot N}.
  \label{eq:gustafson-peff}
\end{equation}
Thus Gustafson can be written in exactly the same total-time form as
Eq.~\eqref{eq:amdahl-time}, except that the effective parallel fraction
is no longer fixed: it becomes \(\tilde{P}(N)\).
The two are classically equivalent in spirit, but under different
normalizations of workload growth.  Amdahl holds the serial--parallel
split fixed as machine scale changes.  Gustafson's key move is to let
that effective ratio vary with scale, so the growing machine is paired
with a correspondingly growing workload.
Both are historically important, but both remain framed in the language of
replication, processor count, and speedup.  Figure~\ref{fig:legacy-laws}
places that language in context.  The present paper keeps the concern with
performance under constraint, but changes the physical question: not how much
speedup replication can buy, but how constrained resources should be allocated
across heterogeneous hardware when the workload itself is changing.

\section{The Modernized Amdahl Model}
\label{sec:model}

The central change in physical description is simple.  Classical
Amdahl-style analysis is written in terms of serial versus parallel
work, processor count, and speedup under replication.  Modern systems
instead face a different design question: how should a constrained
hardware budget be allocated between specialized logic and programmable
compute when the workload itself is changing?

We therefore introduce three variables:
\begin{itemize}
  \item $x \in [0,1)$: the fraction of constrained hardware resource
        allocated to specialized logic;
  \item $R > 1$: the relative efficiency advantage of specialized
        hardware over programmable compute on the bounded portion of the
        workload;
  \item $S \in [0,1]$: the value-scalable fraction of the workload, meaning
        the portion for which additional effective or logical compute
        continues to deliver scaling-law gains over the design interval
        under consideration.
\end{itemize}

The corresponding normalized execution time is
\begin{equation}
  T(x) = \frac{1-S}{1+(R-1)x} + \frac{S}{1-x}.
  \label{eq:tx}
\end{equation}
This setup makes five assumptions explicit.  First, $S$ and $1-S$ are
normalized workload shares over the design interval under consideration.
Second, specialization improves only the bounded share through the
relative efficiency ratio~$R$.  Third, the value-scalable share remains
on the programmable side because that is where continuing efficiency
gains, software innovation, and model adaptation can still be
productively absorbed.  Fourth, for a given chip or cluster, the model
asks how the available substrate is turned into more effective or
logical compute.  Fifth, the hardware trade-off
is reduced to one constrained allocation dimension~$x$, so the model is
deliberately a first-order allocation law rather than a full chip-level
microarchitectural description.
The first term models the effectively bounded portion of the workload,
which can benefit from specialization; the second models the
value-scalable portion, which remains on the programmable side because
it continues to absorb additional effective compute.

Differentiating twice gives
\begin{equation}
  T''(x)
  =
  \frac{2(1-S)(R-1)^2}{\bigl(1+(R-1)x\bigr)^3}
  + \frac{2S}{(1-x)^3}.
\end{equation}
This expression is strictly positive on $[0,1)$, so the objective is
strictly convex and admits a unique global minimizer.  The threshold
condition is then obtained by differentiating once:
\begin{equation}
  T'(x)
  = -\frac{(1-S)(R-1)}{\bigl(1+(R-1)x\bigr)^2}
    + \frac{S}{(1-x)^2}.
  \label{eq:deriv}
\end{equation}
At the origin,
\begin{equation}
  T'(0) = -(1-S)(R-1) + S.
  \label{eq:origin-derivative}
\end{equation}
So $T'(0) < 0$ exactly when
\begin{equation}
  S < 1 - \frac{1}{R}.
  \label{eq:threshold}
\end{equation}
Since the objective is strictly convex, this condition separates the
interior regime from the collapse regime.  If $S \ge 1 - 1/R$, the
unique optimum is the boundary point $x^*=0$.

In the interior regime, solving $T'(x)=0$ yields
\begin{equation}
  x^*
  =
  \frac{\sqrt{\dfrac{(1-S)(R-1)}{S}} - 1}
       {\sqrt{\dfrac{(1-S)(R-1)}{S}} + (R-1)}.
  \label{eq:interior-optimum}
\end{equation}

For a fixed scalable fraction $S$, specialization is justified only if
\begin{equation}
  R_c = \frac{1}{1-S}.
  \label{eq:rc}
\end{equation}
Equivalently, for a fixed efficiency ratio~$R$, specialization
collapses once
\begin{equation}
  S_c = 1 - \frac{1}{R}.
\end{equation}
Within the interior regime, the optimal specialization share decreases
monotonically as the value-scalable fraction rises, which follows
directly from Equation~\eqref{eq:interior-optimum}.

The threshold is finite rather than asymptotic.  For example, when
$S=0.9$, specialization requires at least a $10\times$ relative
efficiency advantage; when $S=0.95$, it requires at least $20\times$.
As the value-scalable fraction rises, the bar for specialization rises
rapidly.

\begin{figure}[htbp]
\centering
\begin{tikzpicture}
\begin{axis}[
  darktheme,
  width=0.82\linewidth,
  height=0.46\linewidth,
  domain=0:0.65,
  samples=300,
  xmin=0, xmax=0.65,
  ymin=0.3, ymax=3.0,
  restrict y to domain=0.2:3.2,
  clip=true,
  xlabel={$x$ \;(specialization fraction)},
  ylabel={$T(x)$ \;(normalized execution time)},
  legend pos=north west,
]

\addplot[color=plotblue]
  {0.8/(1+9*x) + 0.2/(1-x)};
\addlegendentry{$S=0.2$}

\addplot[color=plotcyan]
  {0.5/(1+9*x) + 0.5/(1-x)};
\addlegendentry{$S=0.5$}

\addplot[color=plotgreen]
  {0.2/(1+9*x) + 0.8/(1-x)};
\addlegendentry{$S=0.8$}

\addplot[color=plotred]
  {0.1/(1+9*x) + 0.9/(1-x)};
\addlegendentry{$S=0.9$}

\addplot[color=plotyellow]
  {0.05/(1+9*x) + 0.95/(1-x)};
\addlegendentry{$S=0.95$}

\addplot[
  color=black,
  dashed,
  line width=1.2pt,
  domain=0.01:0.9,
  samples=200,
  variable=\s,
]
({
  ((sqrt(9*(1-\s)/\s) - 1) / (sqrt(9*(1-\s)/\s) + 9))
},
{
  (1-\s)/(1 + 9*((sqrt(9*(1-\s)/\s) - 1) / (sqrt(9*(1-\s)/\s) + 9)))
  + \s/(1 - ((sqrt(9*(1-\s)/\s) - 1) / (sqrt(9*(1-\s)/\s) + 9)))
});
\addlegendentry{optimal locus $x^{*}(S)$}

\draw[dashed, plotblue, thick]
  (axis cs:0.3333,0.3) -- (axis cs:0.3333,0.50);
\node[circle, fill=plotblue, inner sep=2.2pt]
  at (axis cs:0.3333,0.50) {};

\draw[dashed, plotcyan, thick]
  (axis cs:0.1667,0.3) -- (axis cs:0.1667,0.80);
\node[circle, fill=plotcyan, inner sep=2.2pt]
  at (axis cs:0.1667,0.80) {};

\draw[dashed, plotgreen, thick]
  (axis cs:0.0476,0.3) -- (axis cs:0.0476,0.98);
\node[circle, fill=plotgreen, inner sep=2.2pt]
  at (axis cs:0.0476,0.98) {};

\draw[dashed, black, thick]
  (axis cs:0,0.3) -- (axis cs:0,1.0);
\node[circle, fill=black, inner sep=2.5pt]
  at (axis cs:0,1.0) {};
\node[black, font=\scriptsize, anchor=south west]
  at (axis cs:0.01,1.03) {$x^{*}=0$};

\end{axis}
\end{tikzpicture}
\caption{Normalized execution time $T(x)$ versus specialization
  fraction~$x$ for $R=10$ and varying~$S$.  Dashed markers indicate the
  optimal allocation~$x^{*}$.  For low~$S$, the curves are U-shaped and
  specialization is beneficial; as~$S$ approaches $S_c=0.9$, the
  optimum collapses toward the origin.  The dashed black curve traces
  the optimal locus, terminating at the collapse point $x^{*}=0$.  Above
  the threshold ($S=0.95$), the curve is monotonically increasing and no
  investment in dedicated hardware is optimal.}
\label{fig:time-allocation}
\end{figure}
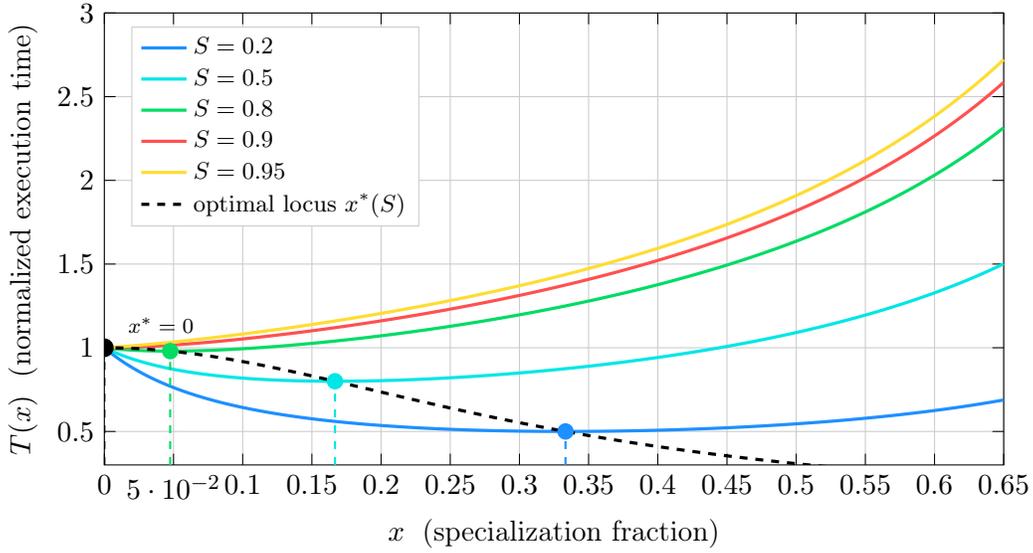

\begin{figure}[htbp]
\centering
\begin{tikzpicture}
\begin{axis}[
  darktheme,
  width=0.82\linewidth,
  height=0.46\linewidth,
  domain=0:0.966,
  samples=300,
  xmin=0, xmax=0.97,
  ymin=1, ymax=30,
  xlabel={$S$ \;(value-scalable fraction)},
  ylabel={$R$ \;(efficiency ratio)},
  legend pos=north west,
]

\addplot[name path=boundary, color=plotcyan, very thick]
  {1/(1-x)};
\addlegendentry{$R_c = \dfrac{1}{1-S}$}

\path[name path=top]    (axis cs:0,30) -- (axis cs:0.97,30);
\path[name path=bottom] (axis cs:0,1) -- (axis cs:0.97,1);

\addplot[fill=plotblue!10, draw=none]
  fill between[of=boundary and top];
\addlegendentry{Specialization region: $x^{*}>0$}
\addplot[fill=plotred!18, draw=none]
  fill between[of=bottom and boundary];
\addlegendentry{Collapse region: $x^{*}=0$}

\end{axis}
\end{tikzpicture}
\caption{Threshold diagram in $(S,\,R)$ space for the optimal
  specialization boundary.  The curve $R_c = 1/(1-S)$ marks the minimum
  efficiency ratio required to justify a nonzero specialized
  allocation.  As~$S$ rises, the efficiency bar for more narrowly
  specialized mechanisms rises with it.  Above the curve, a nonzero
  specialized allocation reduces total execution time; below it, the
  optimal specialized allocation falls to $x^{*}=0$.}
\label{fig:phase-boundary}
\end{figure}

The baseline model treats the specialization advantage~$R$ as a
first-order quantity.  A bandwidth-limited extension is given in
Appendix~\ref{app:bandwidth}; it strengthens the pressure toward
preserving a larger programmable substrate in the interior of the
surface but does not alter the threshold condition itself.

\section{Interpreting the Value-Scalable Fraction}
\label{sec:collapse}

\noindent\textbf{Value-scalable fraction.}
The value-scalable fraction $S$ is the share of normalized workload for
which additional effective or logical compute continues to create
meaningful scaling-law gains over the design interval under
consideration, for a fixed task family, operating regime, and
evaluation criterion.  Depending on the domain, those gains may appear
as improved accuracy, fidelity, capability, robustness, or utility.

This is broader than the classical parallel fraction.  A stage may be
highly parallel yet effectively value-bounded if additional effective
compute mainly raises throughput without materially improving the
delivered result.  Conversely, a stage belongs to $S$ when more
effective compute continues to improve the result itself.  The endpoint cases are excluded only to
avoid degeneracy: $S=1$ gives $x^{*}=0$ immediately, while $S=0$ pushes
the model to the fully specialized endpoint.

In modern AI systems, that distinction is not merely philosophical.
Empirically observed scaling laws~\cite{kaplan2020,lu2026} show that larger
models, richer post-training, and more inference-time compute often
continue to produce measurable gains.  But that growth is not only
about increasing raw scale; it also depends on repeated efficiency
doublings that let a fixed substrate support more effective compute.  In
that sense, $S$ is partly a design choice,
but one anchored in observed value scaling.

\section{Architectural Consequences}
\label{sec:architecture}

The reformulation has several immediate architectural consequences:
\begin{itemize}
  \item $S$ is not the classical parallel fraction; it is the
        value-scalable fraction: the part of the workload for which
        additional effective compute still creates value.
  \item A rise in~$S$ does not imply merely adding more processors; it
        refers to additional effective compute, which may come from
        lower precision, sparsity, software optimization, improved model
        design, or better use of the available substrate.
  \item $R$ is a relative quantity, not a one-way constant; the
        programmable substrate also evolves toward the dominant scalable
        workload, and cannot be treated as standing still while
        specialization improves.
  \item The collapse threshold works in both directions: for a given
        efficiency gap~$R$, there is a critical scalable fraction~$S_c$;
        for a given scalable fraction~$S$, there is a critical required
        efficiency gap~$R_c$.  As value-scalable workload grows, the bar
        for specialization rises with it.
  \item Programmability should not be confused with uniform efficiency
        across all tasks.  It is a flexible but still biased substrate
        for workloads whose value-producing stages keep shifting.
\end{itemize}

As $S$ increases, the contribution of early-stage, fixed-function
computation declines and the dominant portion of execution time shifts
toward dynamically scaling computation.  This leads to
\begin{equation}
  x^{*} \rightarrow 0
\end{equation}

as a design tendency.  The interpretation is not that bottlenecks
disappear, but that bounded stages occupy a shrinking fraction of total
execution time.  Once the scalable portion dominates, the model biases
the optimum away from over-customization and toward more general and
more reconfigurable compute fabrics.

\section{Supporting Examples}
\label{sec:evidence}

In modern AI systems, workload scaling is no longer merely hypothetical.
It has become an empirically grounded structural property of dominant
workloads.  AI scaling laws~\cite{kaplan2020,lu2026} provide empirical evidence
that additional effective or logical compute can continue to improve
model quality in a structured way.  They do not predict how the
corresponding efficiency gains are achieved, but they also place no bar
against them: by making additional effective compute valuable, they keep
rewarding innovations that deliver more logical work on a fixed
substrate.  Pre-training, post-training, and test-time computation have
all become genuine scaling axes, so workload growth is no longer merely
an analytical convenience.  In practice, that scalable share often grows
through software- and model-driven efficiency improvement rather than
through raw hardware increase alone.  In the present model, the
parameter~$S$ captures the fraction of computation associated with these
dynamically scaling workload components.

\subsection{Why Rising $S$ Pulls GPUs Toward Programmability}

\begin{figure}[htbp]
\centering
\begin{minipage}[t]{0.48\linewidth}
\centering
\includegraphics[width=\linewidth]{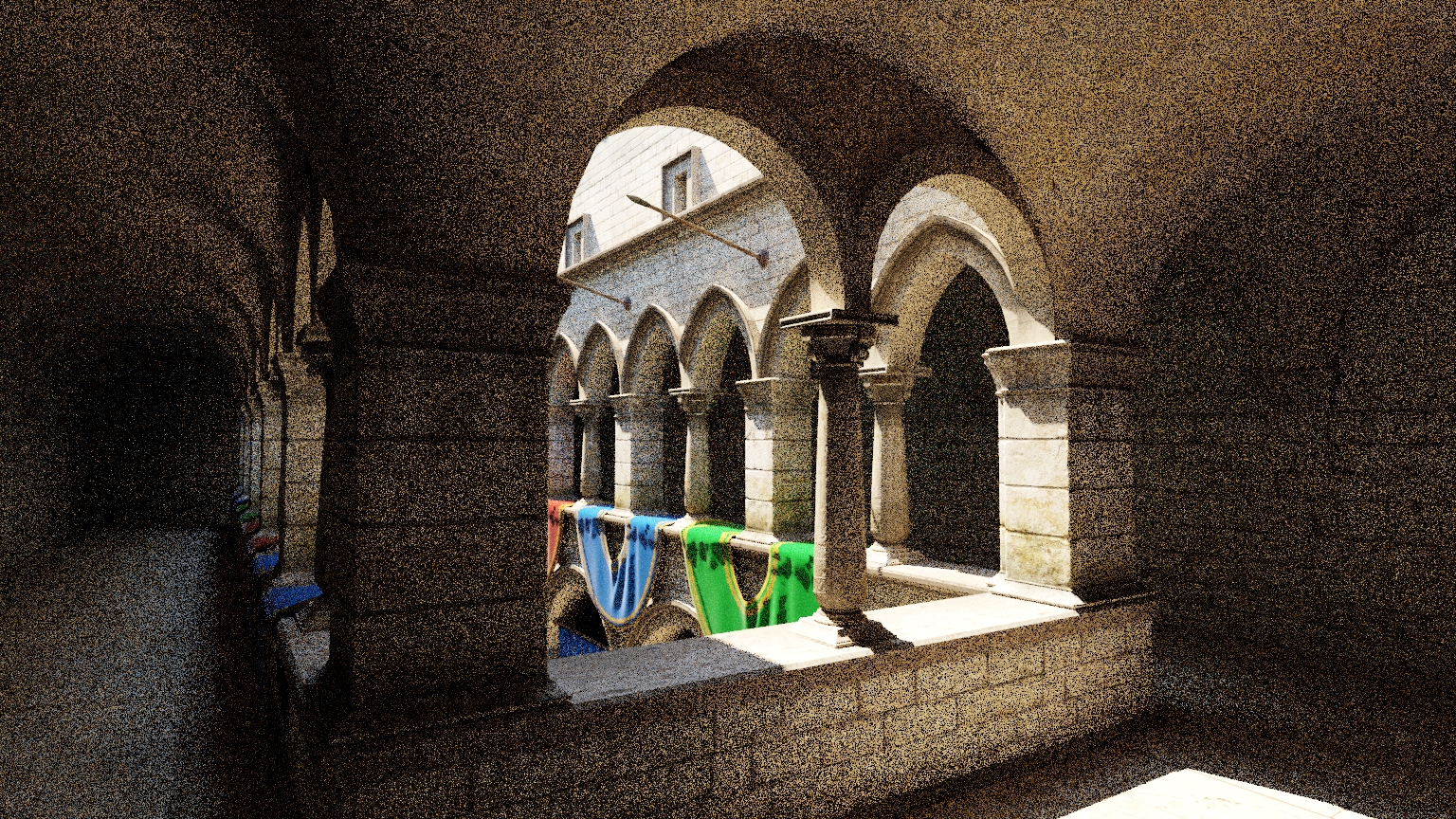}

\small \textbf{Monte Carlo, 16 spp (noisy)}
\end{minipage}
\hfill
\begin{minipage}[t]{0.48\linewidth}
\centering
\includegraphics[width=\linewidth]{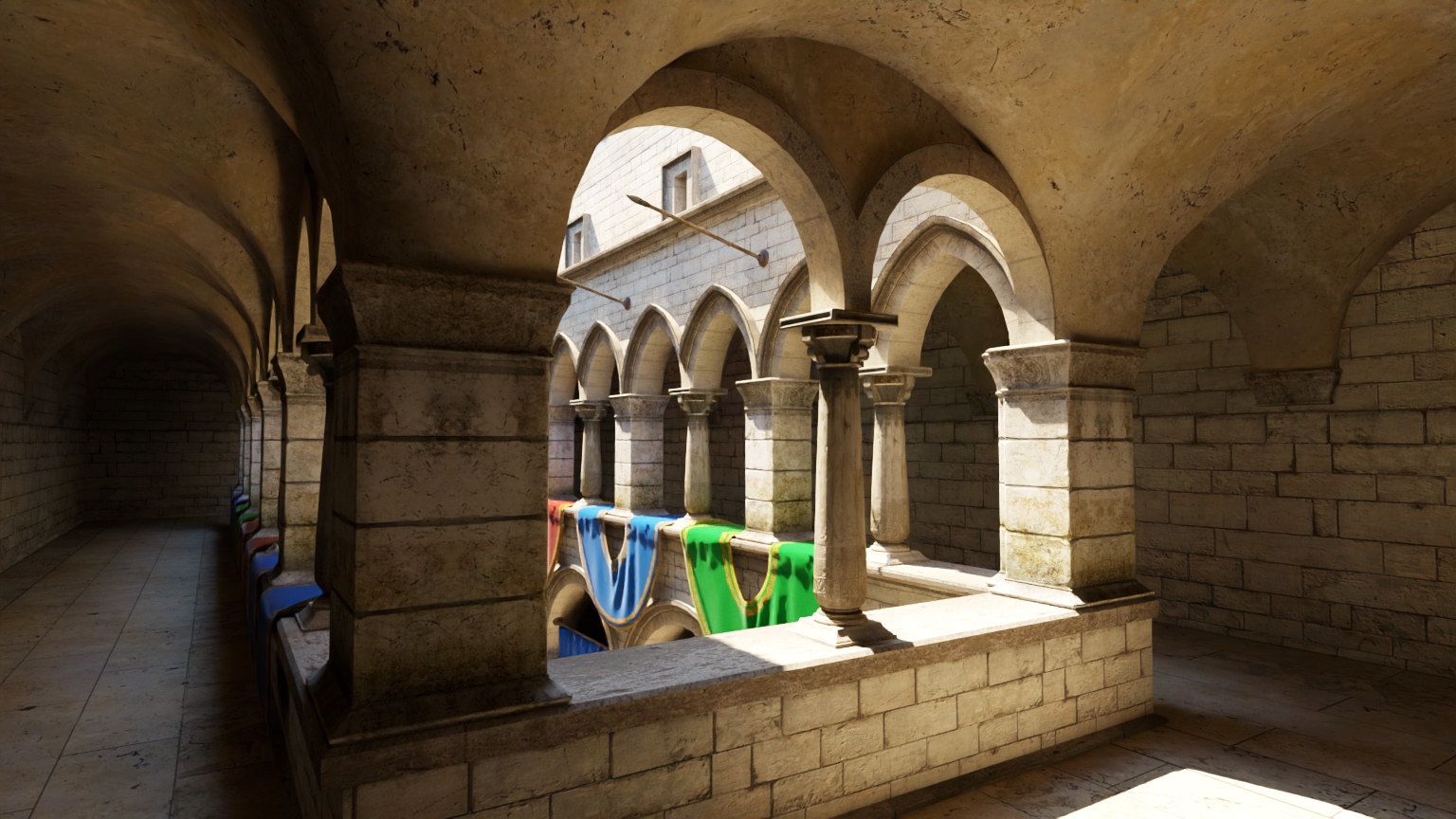}

\small \textbf{Denoised / reconstructed}
\end{minipage}
\caption{Example rendered images illustrating how neural
  denoising and reconstruction shift graphics workload structure.
  Low-sample Monte Carlo rendering provides a noisy acquisition signal,
  while learned denoising recovers useful image quality from that
  input; as reconstruction quality improves, brute-force Monte Carlo
  rendering becomes effectively value-bounded and a larger share of the
  scalable workload shifts into learned post-processing.  The example
  shown is the Crytek Sponza scene at 16 samples per pixel from the
  Intel Open Image Denoise gallery.}
\label{fig:rendering-scalability}
\end{figure}

Rendering makes the distinction concrete.  Classical rasterization is
highly parallel in implementation, but once screen-space resolution and
visibility have been fixed it is not strongly value-scalable: additional
compute adds little new scene information~\cite{akenine2018}.  By
contrast, ray tracing and path tracing without learned reconstruction
remain value-scalable because additional samples continue to improve
fidelity.  Neural denoising and reconstruction change that allocation:
once high perceptual quality can be recovered from low sample counts,
brute-force sampling becomes increasingly value-bounded while neural
inference becomes the stage where marginal compute still adds visible
value~\cite{xiao2020}.  Modern reconstruction systems therefore do not
simply accelerate rendering; they reallocate which stages belong to~$S$
and which become effectively bounded.  Figure~\ref{fig:rendering-scalability}
summarizes this shift.

Across application domains, the growing influence of AI and learned
models is driving shifts such as:
\begin{itemize}
  \item \emph{Reconstruction} replacing direct computation: systems infer a
        full result from partial, noisy, or cheaply sampled inputs
        (e.g.\ denoising, super-resolution, neural codecs, and
        inpainting), not only in rendering but also in signal recovery
        and imaging.
  \item \emph{Late-stage} post-processing dominating execution time: the
        expensive work migrates to stages after a cheaper front end---for
        example neural passes after rasterization, rerankers after
        retrieval, or long/speculative decoding after initial token
        generation.
  \item \emph{Model-driven} output generation: learned models substantially
        produce or shape the delivered artifact, as in language or code
        assistants, speech synthesis, and machine translation.
\end{itemize}

These shifts limit what can be achieved by local optimization of a fixed
pipeline within one vertical.  As the scalable ratio~$S$ rises, more
value-producing work shifts into stages that scale with model capacity
and inference compute, so progress depends less on front-end
specialization and more on the shared scalable path.  That same shift
pulls hardware toward greater programmability.  In graphics, this
predates AI: \emph{deferred shading} already moved more image-quality
generation downstream, and neural denoising, upscaling, and frame
reconstruction intensify that trend by shifting still more work into
learned stages whose cost and capability scale with model complexity and
inference compute~\cite{akenine2018,xiao2020}.  In the language of the
present model, graphics increasingly inherits scaling-law
behavior~\cite{kaplan2020,lu2026}, while front-end visibility and
rasterization look more like bounded acquisition stages.

This is consistent with the long-run trajectory of GPUs:
\begin{itemize}
  \item fixed-function graphics pipelines gave way to shaders---for
        example, hardware transform-and-lighting moved into programmable
        vertex shaders, and fixed texture combiners and per-pixel lighting
        moved into programmable pixel (fragment) shaders,
  \item shaders evolved into unified programmable compute, and
  \item graphics hardware absorbed tensor acceleration, which is now
        increasingly presented as programmable
        matrix machinery rather than rigid frontends.
\end{itemize}

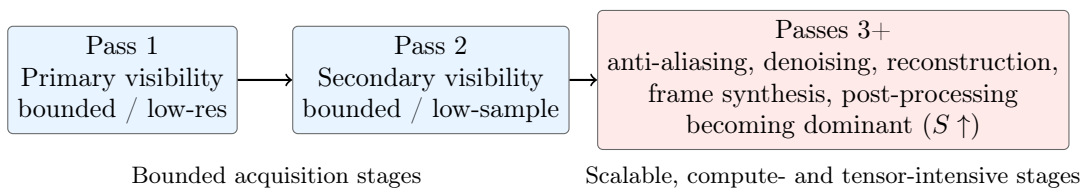
\begin{figure}[htbp]
\centering
\begin{tikzpicture}[x=1cm,y=1cm, font=\small]
  \tikzstyle{classic}=[draw=black!60, rounded corners=2pt, fill=plotblue!10,
    minimum width=2.5cm, minimum height=1.0cm, align=center]
  \tikzstyle{neural}=[draw=black!60, rounded corners=2pt, fill=plotred!12,
    minimum width=5.0cm, minimum height=1.0cm, align=center]

  \node[anchor=west, font=\bfseries] at (0, 1.45)
    {Graphics pipeline under rising $S$};

  \node[classic] (p1) at (1.4,0) {Pass 1\\Primary visibility\\bounded / low-res};
  \node[classic] (p2) at (5.5,0) {Pass 2\\Secondary visibility\\bounded / low-sample};
  \node[neural] (p3) at (10.85,0) {Passes 3+\\anti-aliasing, denoising, reconstruction,\\frame synthesis, post-processing\\becoming dominant ($S \uparrow$)};

  \draw[->, thick] (p1.east) -- (p2.west);
  \draw[->, thick] (p2.east) -- (p3.west);

  \node[anchor=north, align=center, font=\footnotesize] at (3.45,-1.02)
    {Bounded acquisition stages};
  \node[anchor=north, align=center, font=\footnotesize] at (10.85,-1.02)
    {Scalable, compute- and tensor-intensive stages};

\end{tikzpicture}
\caption{Shift of graphics workload structure under rising~$S$.
  In ray-traced or path-traced rendering with learned reconstruction,
  neural denoising and reconstruction compress the classical
  high-sample rendering regime: once useful image quality can be
  recovered from low-resolution or low-sample acquisition, brute-force
  Monte Carlo rendering becomes effectively value-bounded, primary and
  secondary visibility increasingly behave as bounded acquisition
  stages, and passes~3+ absorb a larger share of the scalable workload
  through anti-aliasing, denoising, reconstruction, frame synthesis,
  and related post-processing.  In the extreme limit, many of these
  later passes collapse into a single learned reconstruction stage.}
\label{fig:graphics-pipeline-shift}
\end{figure}

As rendering quality depends more on neural reconstruction than on fixed
multi-pass logic, graphics processors are pulled toward more
programmable, matrix-oriented substrates.  Here, programmability should
be understood in the modern AI-shaped sense, not as generic
instruction-set flexibility alone: the substrate must absorb software
and workload change faster than dedicated silicon can be redesigned.
Figure~\ref{fig:graphics-pipeline-shift} shows that shift at the level
of rendering passes, while Appendix~\ref{app:bandwidth} shows that
bandwidth limits reinforce rather than overturn it.  The same logic
motivates the next question: if scalable work keeps migrating, why do
both GPUs and AI accelerators get pulled toward greater
programmability?

\subsection{Why AI Accelerators Also Become More Programmable}
\label{sec:ai-accelerators}

Within that shared design principle, AI accelerators can still be very
effective when the workload is stable and a
large share of value comes from mechanisms that can be tightly
customized.  But sustained AI scaling creates a different industrial
pressure.  The frontier keeps demanding recurring efficiency gains
through changing model structures, training recipes, dataflows, memory
behavior, and inference-time strategies.  For a fixed chip or cluster,
the issue is therefore repeated efficiency scaling that lets the
available substrate support more effective AI compute.

Once the workload mix described in Section~\ref{sec:evidence} shifts
toward larger values of~$S$, the efficiency advantage of any fixed
mechanism faces a rising bar while scalable work dominates the runtime
budget.  Figure~\ref{fig:phase-boundary} should therefore be read as a
shared design constraint on heterogeneous systems: even below the
boundary, $x^{*}>0$ still implies a mixed system in which programmable
compute remains essential.  That is why successful AI accelerators are
not remaining rigidly fixed-function: like GPUs, they are moving toward
shared tensor, memory, and software abstractions that can absorb
continuing workload change.

\subsection{The TPU as an Early Illustration}
\label{sec:accelerators}

Google's TPUs are an early illustration of the same design principle.
In the sense emphasized by
Hennessy and Patterson, they were introduced as AI
accelerators~\cite{hennessy2019,jouppi2017}, but they did not
specialize around one particular model.  Instead, they elevated dense
tensor computation itself to a broad computational substrate.  In that
sense, they show that
successful specialization in AI often moves upward in abstraction
rather than downward into one narrowly fixed mechanism.

The key point is scaling-law-relevant gain rather than physical possibility.  It is often
possible in principle to scale hardware around a fixed mechanism, but
its architectural payoff declines if software and model evolution move the frontier of
useful computation elsewhere.  In AI, sparsity, routing, cache
compression, quantization, and inference-time system optimization can
reduce cost per token without requiring a corresponding fixed-function
redesign~\cite{fedus2022,deepseek2024}.  In that regime, software
optimization can outrun narrow hardware advantage, so the contribution
of any one dedicated mechanism becomes effectively bounded relative to
the evolving programmable workload.

More broadly, the TPU case suggests that specialization remains
attractive only while the targeted mechanism stays a large and stable
share of value-producing computation.  In the end, this is a shared
design principle between GPUs and future AI accelerators: preserve a
broad programmable substrate rather than over-customizing around one
bounded stage.

\section{Conclusion}
\label{sec:conclusion}

The paper's central result is simple.  Once Amdahl-style analysis is
rewritten in terms of hardware allocation, relative efficiency, and a
value-scalable workload fraction, specialization acquires a finite
collapse threshold.  For a given efficiency ratio~$R$, there is a
critical scalable fraction $S_c = 1 - 1/R$ beyond which the optimal
specialized allocation becomes zero; equivalently, for a given~$S$, the
required efficiency ratio is $R_c = 1/(1-S)$.

The deeper implication is architectural.  In modern AI systems, the
scalable share of work increasingly comes from software- and
model-driven improvements: changing training recipes, inference
strategies, sparsity patterns, routing behavior, cache compression, and
other innovations that can keep creating value on existing hardware.
That is why the value-scalable share tends to stay on the programmable
side of the boundary.  If future gains keep arriving through software
and model evolution, then hardware must remain programmable not only
within one design cycle but across hardware generations, so that the
next round of improvement can still land on the existing substrate
rather than requiring each gain to be hard-wired in advance.
Under that view, scaling remains predictable because the law itself can
stay comparatively stable, while effective cost is beaten down by the
efficiency doublings needed to keep feeding it~\cite{lu2026}.

The claim is therefore not that specialization disappears in every
engineering context.  It is that specialization faces a rising bar when
the value-producing frontier keeps moving.  In that regime,
specialization and programmability are both relative concepts:
specialization means committing budget to mechanisms whose function is
comparatively narrow or bounded, while programmability means preserving
room for continuing software- and model-level innovation.  That is the
shared design principle running through graphics processors, GPUs, and
future AI accelerators alike: preserve enough programmable substrate to
absorb the moving frontier of scalable work rather than over-customizing
around one bounded stage.

Like classical Amdahl's Law, the present reformulation is not a direct
recipe for silicon design.  Its role is conceptual: the classical form
highlights the dominance of the serial bottleneck, whereas the
modernized form highlights the architectural importance of preserving
sufficient programmability under sustained scaling.  A natural extension
is to replace time with energy or energy-delay objectives.  That would
change the allocation objective, but the qualitative conclusion may
still persist: dedicated hardware would still need to exceed a critical
relative efficiency to justify dedicated investment, even if the
quantitative threshold shifts.

\appendix
\section{Bandwidth-Limited Extension}
\label{app:bandwidth}

The baseline model treats the specialization advantage~$R$ as a constant.
That is appropriate for the first-order allocation law, but one obvious
objection is that memory bandwidth can prevent specialized compute from
realizing its full theoretical efficiency.  In Roofline-style
terms~\cite{williams2009}, increasing compute allocation without a
corresponding increase in data supply eventually makes performance
bandwidth-limited.

One simple way to capture this effect is to replace the constant
efficiency ratio by an effective ratio that decays with increasing
specialized allocation:
\begin{equation}
  R_{\mathrm{eff}}(x)
  =
  \frac{R_{\max}}{1 + \gamma R_{\max} x},
  \label{eq:reff}
\end{equation}
where $R_{\max}$ is the peak theoretical efficiency ratio and
$\gamma$ is a memory-friction coefficient summarizing the workload's
arithmetic intensity relative to available memory bandwidth.

The execution-time surface then becomes
\begin{equation}
  T_{\mathrm{mem}}(x)
  =
  \frac{1-S}{1 + \left(R_{\mathrm{eff}}(x)-1\right)x}
  + \frac{S}{1-x}.
  \label{eq:tmem}
\end{equation}

This extension has two useful implications.  First, memory friction
pulls the right-hand side of the surface upward, so the interior optimum
$x^{*}$ moves closer to zero even before the collapse threshold is
reached.  Bandwidth limitations therefore strengthen the pressure toward
programmable hardware.

Second, for the friction model in
Equations~\ref{eq:reff}--\ref{eq:tmem}, the collapse threshold itself is
unchanged.  Evaluating the derivative at the origin gives
\begin{equation}
  \left.\frac{dT_{\mathrm{mem}}}{dx}\right|_{x=0}
  = -(1-S)(R_{\max}-1) + S,
  \label{eq:tmem-deriv}
\end{equation}
which yields the same boundary condition
\begin{equation}
  S_c = 1 - \frac{1}{R_{\max}}.
  \label{eq:tmem-threshold}
\end{equation}

For this bandwidth-friction extension, memory hierarchy effects deform
the interior of the allocation surface without altering the first-order
phase boundary.  The finite collapse remains governed by workload
structure and peak specialization advantage, while bandwidth limitations
flatten the approach to that threshold.



\begin{thebibliography}{10}

\bibitem{akenine2018}
Tomas Akenine-M{\"o}ller, Eric Haines, Naty Hoffman, Angelo Pesce, Micha{\l}
  Iwanicki, and S{\'e}bastien Hillaire.
\newblock {\em Real-Time Rendering}.
\newblock A K Peters/CRC Press, 4 edition, 2018.

\bibitem{amdahl1967}
Gene~M. Amdahl.
\newblock Validity of the single processor approach to achieving large scale
  computing capabilities.
\newblock In {\em Proceedings of the April 18--20, 1967, Spring Joint Computer
  Conference}, AFIPS '67 (Spring), pages 483--485. ACM, 1967.

\bibitem{deepseek2024}
{DeepSeek-AI}.
\newblock Deepseek-v2: A strong, economical, and efficient mixture-of-experts
  language model.
\newblock {\em arXiv preprint arXiv:2405.04434}, 2024.

\bibitem{fedus2022}
William Fedus, Barret Zoph, and Noam Shazeer.
\newblock Switch transformers: Scaling to trillion parameter models with simple
  and efficient sparsity.
\newblock {\em Journal of Machine Learning Research}, 23(120):1--39, 2022.

\bibitem{gustafson1988}
John~L. Gustafson.
\newblock Reevaluating {Amdahl's} law.
\newblock {\em Communications of the ACM}, 31(5):532--533, 1988.

\bibitem{hennessy2019}
John~L. Hennessy and David~A. Patterson.
\newblock A new golden age for computer architecture.
\newblock {\em Communications of the ACM}, 62(2):48--60, 2019.

\bibitem{jouppi2017}
Norman~P. Jouppi, Cliff Young, Nishant Patil, et~al.
\newblock In-datacenter performance analysis of a tensor processing unit.
\newblock In {\em Proceedings of the 44th Annual International Symposium on
  Computer Architecture}, pages 1--12, 2017.

\bibitem{kaplan2020}
Jared Kaplan, Sam McCandlish, Tom Henighan, Tom~B. Brown, Benjamin Chess, Rewon
  Child, Scott Gray, Alec Radford, Jeffrey Wu, and Dario Amodei.
\newblock Scaling laws for neural language models.
\newblock {\em arXiv preprint arXiv:2001.08361}, 2020.

\bibitem{lu2026}
Chien-Ping Lu.
\newblock The unreasonable effectiveness of scaling laws in ai.
\newblock {\em arXiv preprint arXiv:2603.28507}, 2026.

\bibitem{williams2009}
Samuel Williams, Andrew Waterman, and David Patterson.
\newblock Roofline: An insightful visual performance model for multicore
  architectures.
\newblock {\em Communications of the ACM}, 52(4):65--76, 2009.

\bibitem{xiao2020}
Lei Xiao, Salah Nouri, Matt Chapman, Alexander Fix, Douglas Lanman, and Anton
  Kaplanyan.
\newblock Neural supersampling for real-time rendering.
\newblock {\em ACM Transactions on Graphics}, 39(4):142:1--142:12, 2020.

\end{thebibliography}
\end{document}